\documentclass[a4paper,11pt]{article}
\usepackage{pos,amsmath,graphicx,bm}

\ShortTitle{Pion screening mass to $\mathcal{O}(\mu_\ell^2)$ from lattice QCD}
\title{A New Way to Compute the Pseudoscalar Screening Mass at Finite Chemical Potential}

\author*[a]{Prasad Hegde}
\author[b]{Rishabh Thakkar}

\affiliation[a]{Centre for High Energy Physics,\\ Indian Institute of Science,\\
  Bangalore 560012, India.}

\affiliation[b]{Key Laboratory of Quark and Lepton Physics (MOE) \\and Institute of Particle Physics,\\ Central China Normal University,\\
Wuhan 430079, China.}

\emailAdd{prasadhegde@iisc.ac.in}
\emailAdd{rishabh@ccnu.edu.cn}

\abstract{We present a method to calculate the pion screening mass in 2+1-flavor lattice QCD to $\mathcal{O}(\mu^2_\ell)$, where $\mu_\ell$ is the isoscalar chemical potential. Our approach is based on the expression for the free theory pion screening correlator for massless quarks. We use the Taylor expansion method to calculate the screening correlator to $\mathcal{O}(\mu^4_\ell)$. We then extract the $\mathcal{O}(\mu^2_\ell)$ Taylor coefficient of the screening mass from the Taylor coefficients of the correlator, for two temperatures in the range 2~-~3~GeV. Our calculations were done using the Highly Improved Staggered Quark action, and the strange and light quark masses were set respectively to their physical and nearly physical values, corresponding to meson masses $M_{\bar{s}s}=686$ MeV and $M_\pi=160$ MeV.}

\FullConference{The 40th International Symposium on Lattice Field Theory (Lattice 2023)\\
July 31st - August 4th, 2023\\
Fermi National Accelerator Laboratory\\}

\makeatletter
\newsavebox{\@brx}
\newcommand{\llangle}[1][]{\savebox{\@brx}{\(\m@th{#1\langle}\)}%
  \mathopen{\copy\@brx\kern-0.5\wd\@brx\usebox{\@brx}}}
\newcommand{\rrangle}[1][]{\savebox{\@brx}{\(\m@th{#1\rangle}\)}%
  \mathclose{\copy\@brx\kern-0.5\wd\@brx\usebox{\@brx}}}
\makeatother

\newcommand{\m}{\mu_\ell}
\newcommand{\x}{\bm{x}}
\renewcommand{\O}{\mathcal{O}}
\newcommand{\Z}{\mathcal{Z}_\text{QCD}}
\newcommand{\M}{\mathcal{M}}
\newcommand{\tr}{\text{tr}}
\newcommand{\z}{\hat{z}}
\renewcommand{\hm}{\hat{\mu}_\ell}

\begin{document}
\maketitle

\section{Introduction}
\label{sec:introduction}
The properties of the quark-gluon plasma (QGP), which is a new state of nuclear matter that forms at high temperatures, is of great theoretical interest. The QGP has also been created in ultra-relativistic collisions of heavy nuclei at ultra-relativistic energies, thus providing additional motivation for its study. The system created in these experiments has been found to be strongly-coupled,  hence a non-perturbative approach is required for its study~\cite{PHENIX:2004vcz}. For small baryochemical potentials $\mu_B$, such an approach is provided by lattice QCD despite the presence of the sign problem~\cite{Cuteri:2023evl}. At larger $\mu_B$ however, the sign problem becomes much more severe due to which the calculation breaks down~\cite{deForcrand:2009zkb, Aarts:2013naa}.

At very high temperatures and/or densities, the strong coupling constant $\alpha_s$ approaches zero due to asymptotic freedom and hence a perturbative approach might be expected to be valid. Unfortunately, standard finite-temperature QCD perturbation theory breaks down at higher orders due to the infrared divergences of the theory~\cite{Linde:1980ts,Gross:1980br}. Even at lower orders, the series is known to converge poorly. Hence it becomes necessary to resum the perturbation series. Resummed perturbation theory has provided reliable estimates of several thermodynamic observables~\cite{Kajantie:2002wa}. However as the temperature is decreased, the system becomes increasingly non-perturbative and the resummed estimates become unreliable.

Aside from bulk observables such as the pressure or energy density, a second type of observables is the thermal correlation functions of various operators. Meson screening correlators are an important example of this second type of observables. In these proceedings, we will present results for the pion screening correlator obtained at finite temperature and isoscalar chemical potential $\m$ using lattice QCD. To address the sign problem, we shall make use of the method of Taylor expansions~\cite{Gavai:2003mf,Allton:2003vx}. We will also present a method to calculate the pion screening mass to $\O(\m^2)$ from the Taylor expansion of the screening correlator. We had earlier presented some preliminary results at the previous Lattice conference~\cite{Thakkar:2022frk}. A complete discussion of our method and results can be found in our paper~\cite{Thakkar:2023nvr}.

\section{Pion screening correlator at finite $\m$}
\label{sec:scrcorr}
Consider lattice QCD with $N_f=2+1$ flavors of staggered quarks. The partition function $\Z$ at temperature $T$ and isoscalar chemical potential $\m=\mu_u=\mu_d$, $\mu_s=0$, is given by
\begin{equation}
    \Z(T,\m) = \int \mathcal{D}U e^{-S_G(T)} \Delta(T,\m),
\end{equation}
where the integration is over all the gauge links $U$, $S_G(T)$ is the gauge action, and $\Delta(T,\m)$ is the fermion determinant given by
\begin{equation}
    \Delta(T,\m) = \prod_{f=u,d,s} \left[\det M_f(m_f,T,\mu_f)\right]^{1/4}.
\end{equation}

The meson screening correlators $C(z,T,\m)$ are obtained by summing the corresponding two-point correlation functions over all coordinates except the $z$ coordinate viz. 
\begin{align}
 C(z,T,\m) &= \frac{1}{N_\tau N_\sigma^2} \sum_{x,y,\tau}\frac{1}{\Z(T,\m)}\int\mathcal{D}U\,e^{-S_G(T)}\,\Delta(T,\m)\left[\overline{\M}(x,y,z,\tau)\M(0,0,0,0)\right], \notag \\
 &\equiv \frac{1}{N_\tau N_\sigma^2} \sum_{x,y,\tau} \llangle[\big] \overline{\M}(x,y,z,\tau)\M(0,0,0,0) \rrangle[\big],
\label{eq:scrcorr}
\end{align}
where the double angular brackets $\llangle\cdot\rrangle$ denote the expectation value at $\m\neq 0$. 

A typical staggered meson operator is given by $\M(\x) = \sum_{\x'}\phi_{ij}(\x,\x')\bar{\chi}_i(\x)\chi_j(\x')$, where $\x\equiv(x,y,z,\tau)$, $\bar{\chi}_i(\x)$ and $\chi_j(\x')$ are the one-component staggered quarks of flavors $i$ and $j$ respectively, $\x$ and $\x'$ belong to the same unit hypercube, and $\phi_{ij}(\x,\x')$ is a phase factor that depends upon the spin and taste of the staggered meson~\cite{Altmeyer:1992dd}. The staggered action preserves a remnant $U(1)$ symmetry of the full chiral symmetry group of continuum QCD. This symmetry is spontaneously broken by QCD interactions, giving rise to a Goldstone pion. The corresponding meson operator is given by $\M(\x) = \bar{\chi}_u(\x)\chi_d(\x)$ i.e. $\phi_{ud}(\x,\x')=\delta_{\x,\x'}$ for all $\x$~\cite{Cheng:2010fe}. Substituting this into Eq.~\eqref{eq:scrcorr} and carrying out the Wick contractions, we get
\begin{equation}
    C(z,T,\m) = \frac{1}{N_\tau N_\sigma^2} \sum_{x,y,\tau}\llangle[\big]\tr\big[P_u(\x,0,\m) P_d^\dagger(\x,0,-\m)\big] \rrangle[\big],
\label{eq:scrcorr_prop}
\end{equation}
where $P_u(\x,0,\mu_u)$ and $P_d(\x,0,\mu_d)$ are the up and down quark propagators from $0$ to $\x$ respectively.

Owing to the sign problem of lattice QCD, Eq.~\eqref{eq:scrcorr_prop} cannot be calculated directly. Instead, we shall adopt the Taylor expansion approach~\cite{QCD-TARO:2001lhr, Pushkina:2004wa, Nikolaev:2020vll}. Expanding Eq.~\eqref{eq:scrcorr_prop} in a Taylor series in $\hm\equiv \m/T$, we get
\begin{equation}
    C(z,T,\m) = \sum_{k=0}^\infty C_k(z,T)\frac{\hm^k}{k!}.
\end{equation}
The first few Taylor coefficients are given by~\cite{QCD-TARO:2001lhr}
\begin{subequations}
\begin{align}
       C_0(z,T) &= \frac{1}{N_\tau N_\sigma^2}\sum_{x,y,\tau}\langle G \rangle, \\ 
       C_1(z,T) &= \frac{1}{N_\tau N_\sigma^2}\sum_{x,y,\tau}\left\langle G' + G\frac{\Delta'}{\Delta}\right\rangle, \\
    C_2(z,T) &= \frac{1}{N_\tau N_\sigma^2}\sum_{x,y,\tau}\left[\left\langle  {G}'' + 2{G}'\, \frac{ {\Delta}' }{ \Delta }
+G \,\frac{{\Delta}'' }{ \Delta } \right\rangle
-  \left\langle G \right\rangle  \left\langle \frac{ {\Delta}'' }{ \Delta } \right\rangle \right], \quad \text{etc.}
\end{align}
\label{eq:Ck_zero_to_two}
\end{subequations}
where
\begin{equation}
G(\x,\m) \equiv\tr \big[P(\x,0,\m) P^\dagger(\x,0,-\m)\big],
\end{equation}
and the single angular brackets denote expectation values at $\m=0$ viz.
\begin{equation}
\left\langle\O\right\rangle = \frac{1}{\Z(T,0)}\int \mathcal{D}U\,\O \, e^{-S_G(T)} \Delta(T,0).
\end{equation}

\section{Free Theory Correlator}
\label{sec:tay_exp}
The above correlator, Eq.~\eqref{eq:scrcorr}, is known exactly for massless free quarks in the continuum~\cite{Vepsalainen:2007ke}. For $\z\equiv zT \gg 1$, the result is
\begin{equation}
    \frac{C_\text{free}(z,T,\m)}{T^3} = \frac{3e^{-2\pi \z}}{2\z}
    \left[\left(1+\frac{1}{2\pi \z}\right)\cos(2z\m) + \frac{\hm}{\pi}\sin(2z\m)\right] + \O\left(e^{-4\pi\z}\right).
\label{eq:scr_corr_free}
\end{equation}
We see that for $\m=0$, the correlator decays like $C(z) \sim e^{-Mz}/z$ for large $z$, where $M(T,0)=2\pi T$ is the $\m=0$ screening mass. For $\m\neq0$, although the screening mass $M(T,\m)$ is still equal to $2\pi T$, the $\cos(2z\m)$ and $\sin(2z\m)$ factors superimpose an oscillation on the original exponential decay. We can still express $C(z,T,\m)$ as an exponential provided we let the screening mass and the correlator amplitude take complex values i.e.
\begin{align}
    \frac{C_\text{free}(z,T,\m)}{T^3} &= \text{Re}\Big[A(T,\m)e^{-zM(T,\m)}\Big], \notag \\
    &= e^{-zM_R(T,\m)}\Big[A_R(T,\m)\cos(zM_I(T,\m)) + A_I(T,\m)\sin(zM_I(T,\m))\Big], \label{eq:ansatz} \\[0.1cm]
    M(T,\m) &= 2\pi T + 2i\m \equiv M_R(T,\m) + iM_I(T,\m), \label{eq:Mfree} \\[0.1cm]
    A(T,\m) &= \frac{3}{2\z}\left(1+\frac{1}{2\pi\z}\right)\left(1 - i\,\frac{\hm}{\pi}\right) \equiv A_R(T,\m) - iA_I(T,\m). \label{eq:Afree}
\end{align}
By Taylor-expanding Eq.~\eqref{eq:scr_corr_free} w.r.t.~$\hm$, we can obtain the Taylor coefficients of the free theory screening correlator. We shall discuss our results for the Taylor coefficients in the next section. However, to determine the $\m$-corrections to the screening mass, we will also require the following ratios of Taylor coefficients:
\begin{equation}
\Gamma(\z) \equiv \frac{C_2(z,T)}{C_0(z,T)} = - 4\z^2 + \frac{4\z}{\pi} -\frac{2}{\pi^2}+\O(\z^{-1}),
\label{eq:gamma_free}
\end{equation}
and
\begin{equation}
\Sigma(\z) \equiv \frac{C_4(z,T)}{C_0(z,T)} = 16\z^4 - \frac{32\z^3}{\pi} + \frac{16\z^2}{\pi^2}+\O(\z).
\label{eq:sigma_free}
\end{equation}
The exponential factor cancels out in the ratios $\Gamma(\z)$ and $\Sigma(\z)$, which are described by quadratic and quartic polynomials respectively in the large-$\z$ limit.

\section{Free Theory Results}
\label{sec:free_theory_results}
\begin{figure}
    \begin{center}
    \includegraphics[width=.48\linewidth]{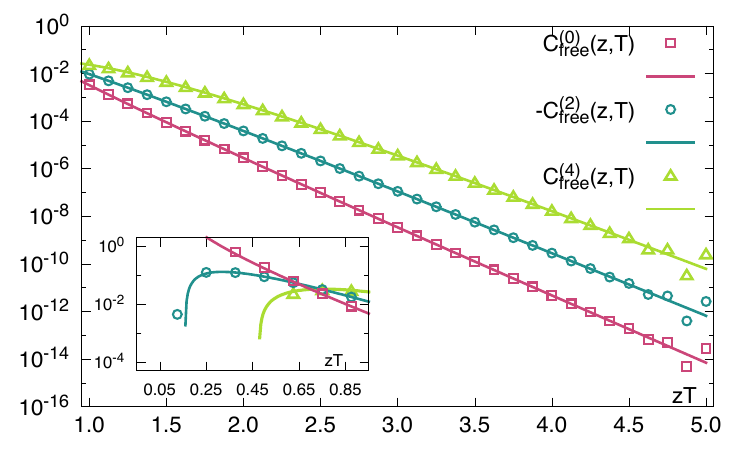}        
    \includegraphics[width=.48\linewidth]{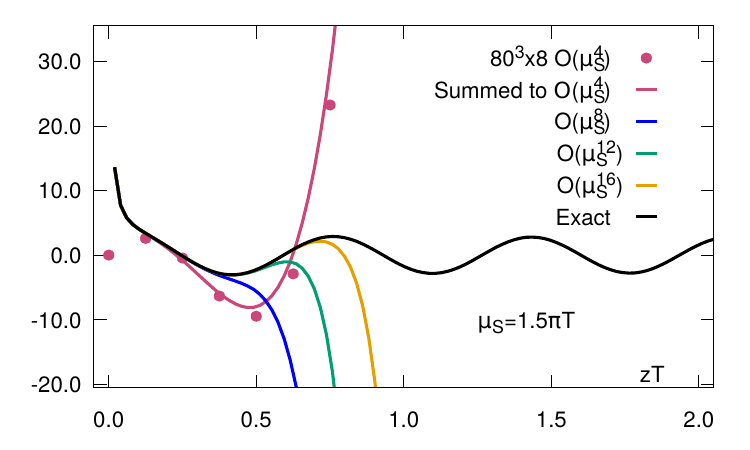}%
    \end{center}
\caption{(Left) Taylor coefficients $C_0$, $C_2$ and $C_4$ (Eq.~\eqref{eq:Ck_zero_to_two}) for the massless free theory correlator. Points are the lattice results while solid lines are the corresponding theoretical predictions (Ref.~\cite{Thakkar:2023nvr}). The main plot shows the results for the range $1 \lesssim \z \leq 5$, while the results for $0 \leq \z \lesssim 1$ are plotted in the inset. (Right) The full correlator amplitude (Eq.~\eqref{eq:amp}), and its Taylor reconstructions  up to various orders (Eq.~\eqref{eq:sum_amp}), plotted for $\hm=1.5\pi$. The lattice data summed up to $\mathcal{O}(\hm^4)$ are also plotted. Figures taken from Ref.~\cite{Thakkar:2023nvr}.}
\label{fig:free_CbyC_isoscalar}
\end{figure}
To verify the above predictions, we calculated the free theory pion screening correlator and its derivatives numerically on an $80^3\times8$ lattice. For the calculation, we used a modified version of the Bielefeld code that was used in the HotQCD collaboration's recent 2+1-flavor meson screening masses calculation~\cite{Bazavov:2019www}. The free theory was simulated by setting all the gauge links to unity, and the HISQ operator was used to calculate the correlators~\cite{Follana:2006rc}. To ensure the convergence of the inverter, it was necessary to work with a small quark mass. However by varying the quark mass, we verified that our results were independent of the quark mass used. Hence, our results are essentially the same as the results for massless quarks.

Our results for the first three non-vanishing Taylor coefficients are presented in Fig.~\ref{fig:free_CbyC_isoscalar} (left). Due to the reflection symmetry of the Dirac operator, the maximum achievable separation between sink and source was $z_\text{max}/a = \frac{1}{2} N_\sigma = 40$. This meant that the maximum possible value for $\z$ was $\z_\text{max}=N_\sigma/2N_\tau=5$.

Although Eq.~\eqref{eq:scr_corr_free} is an asymptotic formula valid for $\z\gg1$, we see that our results agree with its predictions down to $\z\gtrsim0.3$. This is because the neglected terms in that equation are proportional to $e^{-4\pi\z}$, which is only about 2\% of the leading order result at $\z\sim0.3$.

From Fig.~\ref{fig:free_CbyC_isoscalar}, we see that the Taylor coefficients alternate in sign, which is characteristic of the Taylor expansion of an oscillatory function. We can also check this by trying to reconstruct the full correlator amplitude from its Taylor expansions i.e.
\begin{subequations}
\begin{align}
   A_\text{free}(z,T,\m) &\equiv \left(\frac{C_\text{free}}{T^3}\right)\z\,e^{2\pi \z}= \frac{3}{2}    \left[\left(1+\frac{1}{2\pi \z}\right)\cos(2z\m) + \frac{\hm}{\pi}\sin(2z\m)\right],  \label{eq:amp} \\
   &=\sum_{k=0}^\infty \frac{A^{(k)}_\text{free}(z,T)}{k!}\hm^{k}. \label{eq:sum_amp}
\end{align}
\end{subequations}
We plot the exact result for $A_\text{free}$, along with its Taylor reconstructions to different orders, for $\hm=1.5$, in Fig.~\ref{fig:free_CbyC_isoscalar} (right). We see that retaining more terms in the Taylor series allows us to reconstruct the exact correlator up to a greater value of $\z$, beyond which the reconstructed correlator diverges to $\pm\infty$. Alongside the theoretical curves, we also plot its reconstruction as obtained from our lattice results and find that it agrees well with the fourth-order curve. In this way, our calculations allow us to verify the free theory predictions, and this in turn serves as a cross-check of our code.

\section{Screening Mass at Finite $\m$}
\label{sec:scrmass_finite_mu_l}
We assume that at very high temperatures, the screening correlator $C(z,T,\m)$ is described by an equation similar to Eq.~\eqref{eq:ansatz}, with the difference that $M_R$, $M_I$, $A_R$ and $A_I$ are now unknown functions of $T$ and $\m$. Next, we expand $C(z,T,\m)$ in a Taylor series in $\hm$ and construct $\Gamma(\z)$ and $\Sigma(\z)$ out of the Taylor coefficients. We obtain
\begin{align}
\Gamma(z) &= \frac{A_R''}{A_R} 
+ z\left[2\frac{A_I'}{A_R}M_I' - M_R''\right]
- z^2\left(M_I'\right)^2, \notag \\
 &\equiv \alpha_2\z^2+\alpha_1\z+\alpha_0, \label{eq:Gamma}\\
\Sigma(z) &= \frac{A_R''''}{A_R} + z\left[4\frac{A_I'}{A_R}M_I'''+4\frac{A_I'''}{A_R}M_I'-M_R''''-6M_R'' \frac{A_R''}{A_R}\right] \notag\\
&+z^2\left[3M_R''^2-12\frac{A_I'}{A_R}M_I'M_R''-4M_I'M_I'''-6 M_I'^2\frac{A_R''}{A_R}\right]
\notag\\
&+z^3\left[6M_R''M_I'^2-4\frac{A_I'}{A_R}M_I'^3\right]+z^4\left(M_I'\right)^4,
\notag \\
&\equiv\beta_4\z^4+\beta_3\z^3+\beta_2\z^2+\beta_1\z+\beta_0.  \label{eq:Sigma}
\end{align}
The primes denote differentiation w.r.t.~$\hm$ at $\m=0$. It can be shown that $M_R$ and $A_R$ ($M_I$ and $A_I$) must be even (odd) functions of $\m$. This follows from the reality of the screening correlator in Eq.~\eqref{eq:ansatz}. We have therefore set all odd derivatives of $M_R$ and $A_R$ (all even derivatives of $M_I$ and $A_I$) to zero in the above formulas.

We see that $\Gamma(\z)$ and $\Sigma(\z)$ are described by quadratic and quartic polynomials, similar to the free theory case. The coefficients of the polynomials are functions of $M_R''$, $A_I'$, etc. From the coefficients, we see that
\begin{align} && && &&
\hat{M}_I'=\left(-\alpha_2\right)^{1/2}=\beta_4^{1/4} && \text{and} &&
\hat{M}_R''=\frac{1}{4}\left(2\alpha_1-\frac{\beta_3}{\alpha_2}\right). && && &&
\label{eq:mr2}
\end{align}
where $\hat{M}=M/T$. In this way, by fitting the lattice results for $\Gamma(\z)$ and $\Sigma(\z)$ to quadratic and quartic polynomials respectively, we can obtain the leading-order corrections $M_I'\hm$ and $\frac{1}{2}M_R''\hm^2$ to the $\m=0$ screening mass $M(T,0) \equiv M_R(T,0)$.

\section{Finite Temperature Results}
\label{sec:finite_temp_results}
\begin{table}[!h]
	\centering
	\begin{tabular}{ |c|c|c|c|c| } 
		\hline
		$\beta$ & $T$ [GeV]& $N^3_\sigma\times N_\tau$ & configurations \\ 
		\hline
		9.670 & 2.90 & $32^3\times8$ & 12700\\
		      &      & $64^3\times8$ & \phantom{1}6000\\
		9.360 & 2.24 & $64^3\times8$ & \phantom{1}6000\\
		\hline
	\end{tabular}
	\caption{Parameters and statistics for the finite temperature runs.\label{tab:conf_list}}
\end{table}
To obtain $M_R''$ and $M_I'$, we generated 2+1-flavor lattice ensembles for two temperatures viz. $T=2.24$~GeV and $T=2.90$~GeV. The lattices were generated using a Symanzik-improved Wilson action for the gauge fields and the HISQ action for the fermion fields~\cite{Bazavov:2010sb, Bazavov:2011nk}. The temporal extent of the lattices was fixed to $N_\tau = 8$, while the spatial extent was chosen equal to  $N_\sigma = 64$ or 32. Our simulation parameters are listed in Table~\ref{tab:conf_list}. The strange quark mass was set to its physical value, using the updated Line of Constant Physics (LCP) provided in Ref.~\cite{Bazavov:2019www}, while the light quark mass $m_l$ was set equal to $m_s/20$. Further details regarding the runs, as well as regarding the various operators required to calculate $C(z,T,\m)$, can be found in our paper~\cite{Thakkar:2023nvr}.

\begin{figure}[!t]
    \hspace{-0.02\textwidth}%
    \includegraphics[width=.33\linewidth]{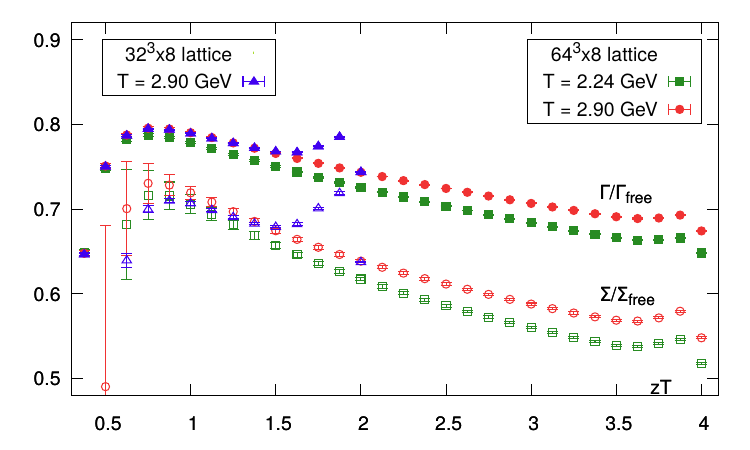}%
    \hspace{0.02\textwidth}%
    \includegraphics[width=.33\linewidth]{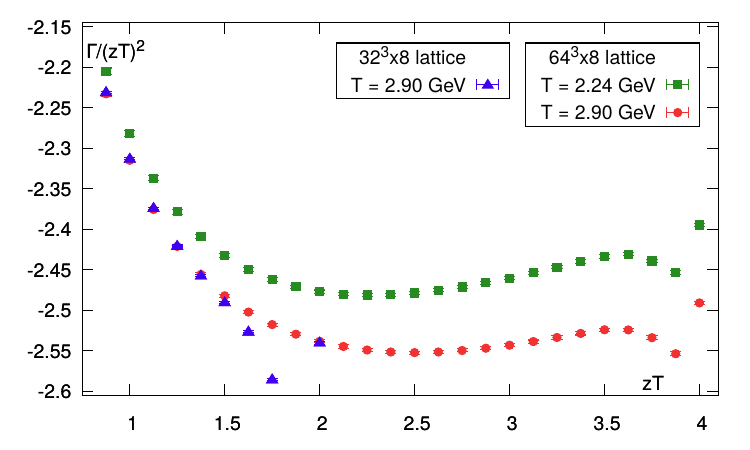}
    \hspace{0.02\textwidth}%
    \includegraphics[width=.33\linewidth]{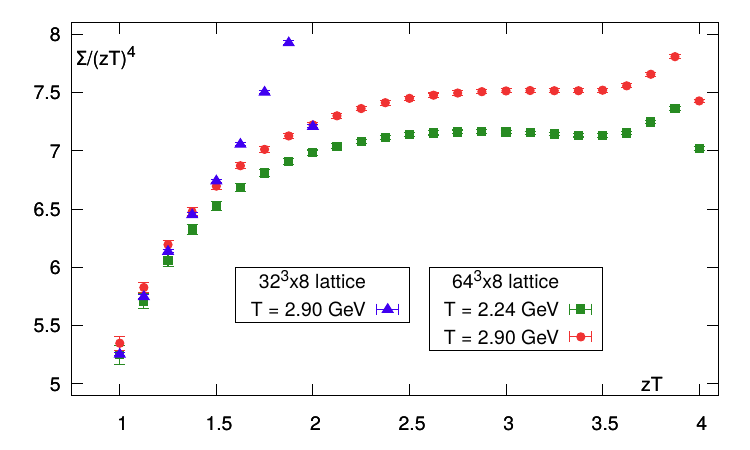}
\caption{(Left) $\Gamma(\z)$ and $\Sigma(\z)$ normalized to the corresponding free theory values. (Middle and Right) $\Gamma(\z)/\z^2$ and $\Sigma(\z)/\z^4$ plotted versus $\z$. All figures from Ref.~\cite{Thakkar:2023nvr}.}
\label{fig:GS_isoscalar}
\end{figure}
In Fig.~\ref{fig:GS_isoscalar}, we present our results for $\Gamma(\z)$ and $\Sigma(\z)$ for all temperatures and volumes. In the left figure, we plot the results after normalizing them to the corresponding free theory values. We see that both $\Gamma(\z)$ and $\Sigma(\z)$ differ from the free theory values by 30-45\% despite the fact that our temperatures are $\sim$15-20 times the chiral crossover temperature~\cite{HotQCD:2018pds}. Our results for the two volumes considered for $T=2.90$ GeV also do not indicate any significant finite-volume effects; however, in all the data sets, both $\Gamma/\Gamma_\text{free}$ and $\Sigma/\Sigma_\text{free}$ curve upwards as $\z_\text{max}=N_\sigma/(2N_\tau)$ is approached, indicating the presence of boundary effects. However, the boundary effects do not seem to affect the point at $\z=\z_\text{max}$.

In the middle and right figures of Fig.~\ref{fig:GS_isoscalar}, we plot $\Gamma(\z)/\z^2$ and $\Sigma(\z)/\z^4$ as functions of $\z$. As discussed in Sec.~\ref{sec:scrmass_finite_mu_l}, we expect $\Gamma$ and $\Sigma$ to be described by 2$^\text{nd}$ and 4$^\text{th}$ degree polynomials in the large-$\z$ limit. Hence, we should expect $\Gamma(\z)/\z^2$ and $\Sigma(\z)/\z^4$ to approach plateaus in the large-$\z$ limit. However, the approach is non-monotonic and we find a minimum (maximum) for $\Gamma(\z)/\z^2$ (for $\Sigma(\z)/\z^4$) respectively.

In fitting our results to Eqs.~\eqref{eq:Gamma} and \eqref{eq:Sigma}, we chose to set $\beta_1=\beta_0=0$ in order to keep the number of fit parameters to a minimum. This reduced the number of fit parameters to three each for $\Gamma(\z)$ and $\Sigma(\z)$. Instead of using all three polynomial coefficients as fit parameters, we used the extremum points $\z_\Gamma$ and $\z_\Sigma$ to reduce the number of fit parameters from three to two by re-expressing $\alpha_0$ and $\beta_2$ in terms of $\alpha_1$ and $\beta_3$. The location of the extrema are given by
\begin{align} && 
\z_\Gamma = -2 \,\frac{\alpha_0}{\alpha_1}\,, &&
\z_\Sigma = -2 \,\frac{\beta_2}{\beta_3}\,. &&
\label{eq:zGamma_zSigma}
\end{align}
Reducing the number of fit parameters from three to two in this way resulted in better fits to the data. $\z_\Gamma$ and $\z_\Sigma$ were located for each jackknife sample using spline fittings. The fits were then carried out for various fit windows $[\z_\text{min},\z_\text{max}]$, with $\z_\text{max}$ fixed to 3.25 and $\z_\text{min}$ varied to obtain a stable result for the fit coefficients. Our procedure yielded good results for some, but not all, of the fit coefficients. Larger lattices will be required in order to get more reliable values of these coefficients, especially at higher temperatures.

Our final results for $M_R''$ and $M_I'$, based on the above procedure, are given in Table~\ref{tab:results}. In the same table, we also give the free theory values of these Taylor coefficients, obtained using Eq.~\eqref{eq:scr_corr_free}. We see that our results differ significantly from the free theory values, despite being obtained for very high temperatures. Understanding the reason for this, as well as improving the accuracy of the present results, will be the aim of future work.

\begin{table}[!ht]
\begin{center}
\begin{tabular}{|c|c|c|c|} \hline
 Temperature & $\hat{M}_R  (\hm=0)$ &$\hat{M}_R''$ & $\hat{M}_I'$\\ \hline
2.24 GeV&
6.337(1)& 0.263(169) & 1.426(5)\\ \hline
2.90 GeV
& 6.352(1) & 0.172(328) & 1.455(6) \\\hline
Free theory& $2\pi\approx6.283$ & 0 & 2 \\ \hline  
\end{tabular}
\end{center}
\caption{Best fit results for $M_R''$ and $M_I'$ for the two temperatures. From Ref.~\cite{Thakkar:2023nvr}.}
\label{tab:results}
\end{table}

\section{Conclusions}
\label{sec:conclusions}
In these proceedings, we discussed a way to obtain the pion screening mass at finite isoscalar chemical potential $\m$. Our approach was based on the known expression for the massless free theory pion screening correlator~\cite{Vepsalainen:2007ke}. Our procedure treated the screening mass and correlator amplitude as complex quantities for $\m\ne0$. While the real part $M_R$ of the screening mass $M=M_R+iM_I$ causes the screening correlator to decay exponentially with increasing $z$, the imaginary part $M_I$ introduces a periodic variation in $z$ in addition to the exponential decay.

Due to the sign problem, the screening correlator cannot be calculated directly for $\m\neq0$. Instead, we expanded the correlator in a Taylor series to $\O\left(\hm^4\right)$~\cite{QCD-TARO:2001lhr, Pushkina:2004wa, Nikolaev:2020vll}. We showed that the ratios $C_2/C_0$ and $C_4/C_0$ of the Taylor coefficients are described by 2$^\text{nd}$ and 4$^\text{th}$ degree polynomials in $\z\equiv zT$ respectively. We showed how to extract the lowest-order Taylor coefficients $M_I'$ and $M_R''$ of the screening mass from the coefficients of these polynomials. Despite working at high temperature ($T\sim 2$ - 3~GeV), our results for the Taylor coefficients differ significantly from the free theory values. It is possible that the approach to the free theory is very slow~\cite{DallaBrida:2021ddx}. 

A complementary approach to calculating the screening correlator was presented in Refs.~\cite{Lowdon:2022xcl} and \cite{Bala:2023iqu} and is based on a generalization of the K\"all\'en-Lehmann spectral representation of the $T=0$ correlators. Hence, the approach is applicable at temperatures below and just above the chiral crossover temperature $T_{pc}$. Since our approach was based on the free theory, we expect it to be valid at temperatures well above $T_{pc}$. Our results seem to indicate that the QGP remains strongly coupled even at these temperatures. A comparison between these results and resummed perturbation theory calculations of the screening correlator and screening mass may help to better understand the approach to the perturbative and free theory limits.

\bibliographystyle{utphys}
\providecommand{\href}[2]{#2}\begingroup\raggedright\endgroup

\end{document}